\documentclass[12pt]{article}
\usepackage{a4wide,amssymb,cite}
\pdfoutput=1

\usepackage{a4wide,amssymb,graphicx}
\usepackage{epsfig}
\usepackage[usenames,dvipsnames]{color}
\usepackage{slashed}
\usepackage{amssymb,cite,graphicx}
\usepackage{slashed}
\usepackage{amsmath,bm,bbm}
\usepackage{amsfonts}
\usepackage[titletoc,title]{appendix}
\usepackage[small]{caption}
\usepackage[margin=1in]{geometry}
\usepackage[multiple]{footmisc}
\usepackage{mathtools}
\usepackage{slashed}
\usepackage[nottoc]{tocbibind}
\usepackage{xcolor}

\newcommand{\be}{\begin{equation}}
\newcommand{\ee}{\end{equation}}
\newcommand{\bea}{\begin{eqnarray}}
\newcommand{\eea}{\end{eqnarray}}

\def\circa#1{\,\raise.3ex\hbox{$#1$\kern-.75em\lower1ex\hbox{$\sim$}}\,}

\begin{document}

\begin{titlepage}

\rightline{CERN-PH-TH/2019-196}

\begin{centering}
\vspace{1cm}
{\Large {\bf Relaxation of Higgs mass and cosmological constant  \\ \vspace{0.2cm}  with four-form fluxes and reheating}} \\

\vspace{1.5cm}

{\bf Hyun Min Lee}
\vspace{.5cm}

{\it Department of Physics, Chung-Ang University, Seoul 06974, Korea.} 
\\ \vspace{0.2cm}
{\it CERN, Theory department, 1211 Geneva 23, Switzerland. }

(Email: hminlee@cau.ac.kr)

\end{centering}
\vspace{2cm}

\begin{abstract}
\noindent
We consider the most general effective action for the four-form fluxes in the Standard Model coupled to gravity. The Higgs mass parameter can be relaxed to a correct value due to the four-form coupling to the Higgs field and it stops changing due to an extremely suppressed transition probability from the observed cosmological constant to AdS space. We first introduce a non-minimal four-form coupling to gravity and discuss the role of a new scalar field as the inflaton and the conditions for a successful reheating at the end of relaxation.

\end{abstract}

\vspace{3cm}
%


\end{titlepage}

\section{Introduction}

The cosmological constant problem is a notoriously difficult problem in particle physics and cosmology, because there is no working symmetry to protect the cosmological constant from being large. This had led to the early no-go theorem for the cosmological constant problem \cite{cc-review}.  

The four-form flux provides an undetermined constant \cite{duff,witten,cc1,cc2}, enabling the cosmological constant to vary towards a small value.  The probability with the Euclidean action \cite{early} may prefer a small cosmological constant among the distribution of values with different flux parameters.
Although the gauge field corresponding to the four-form flux is not dynamical in 4D, the four-form flux can be changed in the process of creating membranes \cite{membrane}. In this case, the tunneling probability between two configurations with cosmological constants differing by one unit can be defined \cite{tunneling}. 

The four-form fluxes have been used to address the hierarchy problem \cite{hierarchy,Higgscan}, inflation \cite{inflation}, quintessence \cite{quint}, strong CP problem \cite{strongCP}, etc. 
As the gauge field for the four-form flux is dynamical in 5D, it was used to source the warped metric with flat space independent of brane and bulk cosmological constants, known as the self-tuning solutions \cite{self-tuning}.  There was also an interesting novel idea for the cosmological relaxation of the Higgs mass with an axion-like scalar field \cite{relaxation}.

Recently, there is an interesting proposal for relaxing the cosmological constant and the Higgs mass parameter to observed values by the same four-form fluxes \cite{Giudice,Kaloper}.  The key ingredient of the proposal is that there is a dimensionless coupling between the four-form flux and the Higgs field, and the flux parameter takes a weak-scale value to relax the Higgs mass parameter to a correct value. Although there is a need of anthropic argument for the cosmological constant \cite{anthropic}, the tunneling probability between two configurations with different cosmological constants can judge when the flux parameter stops changing. The important issue is then how a non-empty Universe is guaranteed by reheating dynamics at the end of relaxation.

In this work, we consider the most general couplings for the four-form fluxes in 4D. These include another dimensionless non-minimal four-form coupling to gravity in addition to the four-form coupling to the Higgs field. 
The non-minimal four-form coupling to gravity gives rise to an $R^2$ term with negative coefficient, which corresponds to a dynamical scalar field with tachyonic mass. We cure the tachyonic instability with an extra positive $R^2$ term from the beginning and discuss the role of the new dynamical scalar field for inflation and reheating dynamics.

The paper is organized as follows.
We begin with an overview on the model containing the four-form flux in the SM minimally coupled to gravity. Then, we review the relaxation mechanism with the four-form flux for solving the hierarchy problem. Next we give the detailed discussion on the Einstein-frame action in a dual tensor-scalar gravity and explain how inflation/reheating takes place and determine the reheating temperature.

\section{The Model}

We consider a three-index anti-symmetric tensor field $A_{\nu\rho\sigma}$ and its four-form field strength   $F_{\mu\nu\rho\sigma}=4\, \partial_{[\mu} A_{\nu\rho\sigma]}$.
Then, the most general Lagrangian with four-form field couplings in the SM are composed of various terms as follows,
\bea
{\cal L} = {\cal L}_0 +{\cal L}_{\rm int}+ {\cal L}_S +{\cal L}_L+ {\cal L}_{\rm memb} \label{full}
\eea
with
\bea
 {\cal L}_0 &=&  \sqrt{-g} \Big[\frac{1}{2}R +\frac{1}{2} \zeta^2 R^2 -\Lambda -\frac{1}{48} F_{\mu\nu\rho\sigma} F^{\mu\nu\rho\sigma} -  |D_\mu H|^2-V(H)\Big], \label{L0} \\
 {\cal L}_{\rm int} &=& \frac{1}{24} \,\epsilon^{\mu\nu\rho\sigma} F_{\mu\nu\rho\sigma} \,(-c_1 R +c_2 |H|^2),  \label{Lagint} \\
 {\cal L}_S &=&\frac{1}{6}\partial_\mu \bigg[\Big( \sqrt{-g}\,  F^{\mu\nu\rho\sigma} + \epsilon^{\mu\nu\rho\sigma} (c_1 R -c_2 |H|^2) \Big)A_{\nu\rho\sigma} \bigg],  \\
 {\cal L}_L &=& \frac{q}{24}\, \epsilon^{\mu\nu\rho\sigma} \Big( F_{\mu\nu\rho\sigma}- 4\, \partial_{[\mu} A_{\nu\rho\sigma]} \Big),  \label{LL} \\
 {\cal L}_{\rm memb}&=& \frac{e}{6} \int d^3\xi\,  \delta^4(x-x(\xi))\, A_{\nu\rho\sigma} \frac{\partial x^\nu}{\partial \xi^a} \frac{\partial x^\rho}{\partial \xi^b} \frac{\partial x^\sigma}{\partial \xi^c} \,\epsilon^{abc}-T\int d^3\xi\,  \delta^4(x-x(\xi)) \sqrt{-g^{(3)}}. 
\eea
Here, the Higgs potential in the SM is given by
\bea
V(H) = -M^2 |H|^2 +\lambda |H|^4. 
\eea
In the interaction Lagrangian $ {\cal L}_{\rm int} $ in eq.~(\ref{Lagint}), $c_1, c_2$ are dimensionless parameters, both of which are taken to be positive in the later discussion. The four-form coupling to the Higgs $c_2$ was introduced before in the literature \cite{hierarchy,Giudice,Kaloper}, but the non-minimal four-form coupling to gravity $c_1$ is introduced here for the first time.  
We note that ${\cal L}_S$ is the surface term necessary for the well-defined variation of the action with the anti-symmetric tensor field \cite{cc2}, and $q$ in ${\cal L}_L$ (in eq.~(\ref{LL})) is the Lagrange multiplier, and $ {\cal L}_{\rm memb}$ is the membrane action coupled to  $A_{\nu\rho\sigma}$ with membrane charge $e$, and the membrane tension can be also introduced by $T$ with $g^{(3)}$ being the determinant of the induced metric on the membrane.
Here, $\xi^a$ are the membrane coordinates, $x(\xi)$ are the embedding coordinates in spacetime and $\epsilon^{abc}$ is the volume form for the membrane.
We also note that the $R^2$ term in eq.~(\ref{L0}) is introduced to ensure the stability of the non-minimal four-form coupling to gravity\footnote{We note that the most general Lagrangian in the quadratic gravity contains $R_{\mu\nu\rho\sigma}R^{\mu\nu\rho\sigma}$ and a Gauss-Bonnet term. The latter term does not affect our discussion because it is a topological invariant, while the former term would induce a spin-2 ghost particle \cite{stelle}. Moreover, the spin-2 ghost does not change the dynamics of the Higgs-four-form coupling, but it could render the gravitational theory inconsistent at quantum level. But, we assume that the spin-2 ghost is decoupled in the effective theory of gravity and focus on the impact of the $R^2$ term in the later discussion. }, as will be discussed later.  

Then, following the strategy in Ref.~\cite{inflation,quint}, we derive  the equation of motion for $F_{\mu\nu\rho\sigma}$ as follows,
\bea
F^{\mu\nu\rho\sigma}=\frac{1}{\sqrt{-g}}\, \epsilon^{\mu\nu\rho\sigma} \Big( -c_1 R + c_2 |H|^2+q\Big),
\eea
and integrate out $F_{\mu\nu\rho\sigma}$. As a result, we obtain the full Lagrangian (\ref{full}) as
\bea
{\cal L} &=&\sqrt{-g} \Big[\frac{1}{2}R+\frac{1}{2} \zeta^2 R^2 -\Lambda-  |D_\mu H|^2 +M^2 |H|^2 -\lambda |H|^4-\frac{1}{2} (-c_1 R+ c_2 |H|^2+q)^2  \Big] \nonumber \\
&&+ \frac{1}{6}\epsilon^{\mu\nu\rho\sigma} \partial_\mu q A_{\nu\rho\sigma} +\frac{e}{6} \int d^3\xi \, \delta^4(x-x(\xi))\, A_{\nu\rho\sigma} \frac{\partial x^\nu}{\partial \xi^a} \frac{\partial x^\rho}{\partial \xi^b} \frac{\partial x^\sigma}{\partial \xi^c} \epsilon^{abc}. \label{Lagfull}
\eea 
As a result, the equation of motion for $A_{\nu\rho\sigma}$ makes the four-form flux $q$  dynamical, according to
\bea
\epsilon^{\mu\nu\rho\sigma} \partial_\mu q= -e\int d^3\xi \, \delta^4(x-x(\xi))\, \frac{\partial x^\nu}{\partial \xi^a} \frac{\partial x^\rho}{\partial \xi^b} \frac{\partial x^\sigma}{\partial \xi^c} \epsilon^{abc}.
\eea
The flux parameter $q$ is quantized in units of $e$ as $q=e\,n$ with $n$ being integer. 
Whenever we nucleate a membrane, we can decrease the flux parameter by one unit  such that both the Higgs mass and the cosmological constant can be relaxed into observed values in the end.

\section{Dynamical relaxation with four-form fluxes} 

From the result in eq.~(\ref{Lagfull}) apart from the second line, we collect the relevant terms in the following form,
\bea
{\cal L} = \sqrt{-g} \bigg[ \frac{1}{2} f(H,q) R +\frac{1}{2}(\zeta^2-c^2_1) R^2-  |D_\mu H|^2 +M^2_{\rm eff} |H|^2 -\Big(\lambda+\frac{1}{2}c^2_2\Big)  |H|^4 -\Lambda_{\rm eff}   \bigg]  \label{Lagfull2}
\eea 
where
\bea
 f(H,q) &=& 1 + c_1 (c_2 |H|^2+q), \\
  M^2_{\rm eff}(q) &=& M^2 - c_2\, q, \\
  \Lambda_{\rm eff} (q) &=& \Lambda + \frac{1}{2}\, q^2. 
\eea
Then, we find that the Higgs mass parameter and the cosmological constant as well as the Planck mass are variable by the same quantity, the flux parameter $q$. Whenever the membrane nucleation occurs, we can reduce the flux parameter and scan the effective parameters. 
It is interesting to notice that there is an $R^2$ term with negative coefficient proportional to the non-minimal four-form coupling in the original Lagrangian (\ref{Lagint}). Thus, we had to include an $R^2$ term from the beginning to compensate the negative term for stability. 
The correction to the Higgs quartic coupling is independent of the flux parameter so it is absorbed by the tree-level value. 

The membrane is located at the boundary between two consecutive dS space configurations that are defined by the flux parameters and differ by one unit. Then, it is argued the tunneling probability between those configurations is given \cite{tunneling} by
\bea
{\cal P}(n+1\rightarrow n) \approx {\rm exp} \left( -\frac{24\pi^2M^4_P}{\Lambda_{n+1}}\right) \label{probable}
\eea
when $\Lambda_{n+1}\ll T^2/M^2_P$ where $T$ is the membrane tension. 
Therefore, the probability of changing the flux parameter by one unit becomes large in the early stage of the nucleation, but it becomes extremely suppressed at the last stage, making the Universe entering in a metastable state with a small cosmological constant \cite{tunneling,membrane,Giudice,Kaloper}.

In addition to the relaxation of the cosmological constant with four-form fluxes, the Higgs mass parameter is also scanned at the same time. 
For $q>q_c$ with $q_c\equiv M^2/c_2$, the Higgs mass parameter $M^2_{\rm eff}<0$, so electroweak symmetry is unbroken, whereas for $q<q_c$, we are in the broken phase.  For $c_2={\cal O}(1)$ and the membrane charge $e$ of electroweak scale,  we can explain the observed Higgs mass parameter once the flux change stops at $q=q_c-e$ by the previous argument for the tunneling probability \cite{Giudice,Kaloper}.
For $\Lambda<0$, we can cancel a large cosmological constant by the contribution from the same flux parameter until $\Lambda_{\rm eff}$ takes the observed value at $q=q_c-e$, but we need to reply on an anthropic argument for that with $e$ being of order weak scale \cite{anthropic}.

We remark the tunneling rate with membrane nucleation in more detail, in particular, in the last stage of the four-form scanning. 
The tunneling rate from the last dS phase to the true vacuum depends on the bounce action $B$ for the instanton solution with radius ${\bar r}_0$ \cite{coleman,tunneling}, given in the following,
\bea
\gamma\equiv {\bar r}^{-4}_0\, e^{-B} \label{rate}
\eea
where the bounce action is given by
\bea
B=\frac{27\pi^2}{2} \, \frac{T^4}{(\Delta \Lambda)^3}\,\left( 1+\frac{1}{4} r^2_0 H^2\right)^{-2},
 \label{bounce}
\eea
with $r_0=\frac{3T}{\Delta \Lambda}$ being the instanton radius in the absence of gravity,
and the instanton radius ${\bar r}_0$ and the dS radius $H^{-1}$ are given, respectively, by 
\bea
{\bar r}_0&=&\frac{r_0}{1+\frac{1}{4} r^2_0 H^2}, \label{iradius}\\
H^{-1} &=& \frac{\sqrt{3} M_P}{\sqrt{\Delta\Lambda}}.
\eea
Here, $\Delta\Lambda$ is the change of the cosmological constant due to the last tunneling, which is given by $\Delta\Lambda\simeq e q_c$ for $e\ll q_c$ after the last membrane nucleation. 
The gravitational corrections appear due to the curvature of the dS phase outside the membrane, suppressing both the bounce action and the instanton radius. 

We note that when $r_0< 2H^{-1}$, which corresponds to $\frac{T^2}{M^2_P}<\frac{4}{3} \Delta\Lambda$, we can ignore the curvature of the dS spacetime and it is enough to consider the  membrane tension and the four-form action for calculating the bounce action. In this case, the tunneling rate becomes $\gamma\simeq r^{-4}_0\, e^{-B}$ with $B\simeq\frac{27\pi^2}{2} \, \frac{T^4}{(\Delta \Lambda)^3}$, from eq.~(\ref{rate}) with eqs.~(\ref{bounce}) and (\ref{iradius}). 
On the other hand, for $r_0\gtrsim 2 H^{-1}$,  the bounce action is dominated by the curvature of the dS space. This is the case shown in eq.~(\ref{probable}), for which the tunneling rate becomes $\gamma\simeq r^{-4}_0 \Big(\frac{r_0 H}{2} \Big)^8\, e^{-B}$ with $B\simeq \frac{24\pi^2 M^4_P}{\Delta\Lambda}$, similarly from eq.~(\ref{rate}) with eqs.~(\ref{bounce}) and (\ref{iradius}). 

The last dS phase at $q=q_c$ becomes unstable within the Hubble volume when $\gamma> H^4$.
In the case with $r_0< 2H^{-1}$ and $T=M^3_*$, we can obtain the condition on the brane tension for $\gamma> H^4$, as follows,
\bea
M_*< \frac{1}{1.85^{1/12}}\, (\Delta \Lambda)^{1/4}=\frac{1}{1.85^{1/12}}\, (eq_c)^{1/4}.  \label{instable}
\eea
Therefore, for $q_c\sim M^2_P$ and $e\sim (100\,{\rm GeV})^2$, the above instability bound becomes $M_*< 10^{10}\,{\rm GeV}$, which is consistent with the negligible gravity for  $\frac{T^2}{M^2_P}<\frac{4}{3} \Delta\Lambda$, that is, $M_*<4\times 10^{12}\,{\rm GeV}$. 
On the other hand, if the above instability bound (\ref{instable}) is not satisfied, we have $\gamma<H^4$, even though the gravitational corrections become important and help suppress the bounce action, independent of the brane tension with $T>4\times 10^{12}\,{\rm GeV}$. Then, the last dS phase takes a long time before it decays, thus there would be a prolonged stage of the last dS phase.   

In the later discussion on reheating, depending on the brane tension, we divide our discussion into two cases in the next section, namely, reheating the Universe during or after the last membrane nucleation.

\section{Four-form non-minimal couplings and effective theory}

In this section, we discuss the implications of the four-form couplings for the reheating of the Universe.  This is an important ingredient for the non-empty Universe at the end of relaxation. 

We first consider a dual description of the $R^2$ term  in eq.~({\ref{Lagfull2}})  in terms of a real scalar field $\chi$ by
\bea
\frac{1}{2}(\zeta^2-c^2_1) R^2 \longrightarrow \sqrt{\zeta^2-c^2_1}\, \chi R - \frac{1}{2} \chi^2.
\eea
Then, the Lagrangian ({\ref{Lagfull2}}) becomes 
\bea
{\cal L} = \sqrt{-g} \bigg[ \frac{1}{2}\,\Omega(H,\chi,q) R  -  |D_\mu H|^2 +M^2_{\rm eff} |H|^2 -\Big(\lambda+\frac{1}{2}c^2_2\Big)  |H|^4 -\Lambda_{\rm eff} -\frac{1}{2} \chi^2 \bigg]  \label{Lagfull3}
\eea
with
\bea
\Omega(H,\chi,q)=1 + c_1 \Big(c_2 |H|^2+q\Big)+\sqrt{\zeta^2-c^2_1}\, \chi. 
\eea
Furthermore, making the field redefinition by
\bea
\sigma= c_2 |H|^2+q+\frac{\sqrt{\zeta^2-c^2_1}}{c_1}\, \chi,
\eea
we get $\Omega=1 + c_1\sigma$ and  rewrite eq.~({\ref{Lagfull3}})  as
\bea
{\cal L} = \sqrt{-g} \bigg[ \frac{1}{2}\,(1+c_1\sigma ) R  -  |D_\mu H|^2-V(H,\sigma,q)  \bigg] \label{Lagfinal}
\eea
with
\bea
V(H,\sigma,q) = -M^2_{\rm eff} |H|^2 +\Big(\lambda+\frac{1}{2}c^2_2\Big)  |H|^4 +\Lambda_{\rm eff} +\frac{1}{2}\,\frac{c^2_1}{\zeta^2-c^2_1} \Big(\sigma-c_2 |H|^2-q \Big)^2.
\eea
We remark that for $\zeta^2>c^2_1$, the potential for a new scalar field $\sigma$ is bounded from below, so the stability of the potential is ensured even in the presence of the non-minimal four-form coupling to gravity.  For $\zeta^2<c^2_1$, the potential is unbounded from below, so we would need a higher dimensional term for the sigma field to stabilize the potential. 

Due to the field-dependent Einstein term in  eq.~(\ref{Lagfinal}), we make a Weyl scaling of the metric by $g_{\mu\nu}=g^E_{\mu\nu}/\Omega$ and get the Einstein frame Lagrangian as follows,
\bea
{\cal L}_E 
=\sqrt{-g_E} \bigg[ \frac{1}{2} R(g_E) -\frac{3}{4}\,c^2_1\,\Omega^{-2}\,(\partial_\mu\sigma)^2 - \frac{1}{\Omega}\, |D_\mu H|^2- \frac{V(H,\sigma,q)}{\Omega^2} \bigg]. 
\eea

 For $|c_1\sigma|\lesssim 1$, we can make the sigma field kinetic term canonically normalized  by ${\bar\sigma}=\sqrt{\frac{3}{2}}\, c_1 \sigma$ and get the Einstein-frame Lagrangian as 
\bea
{\cal L}_E 
&\approx& \sqrt{-g_E} \bigg[ \frac{1}{2} R(g_E) -\frac{1}{2} (\partial_\mu{\bar\sigma})^2 - |D_\mu H|^2-V(H,{\bar\sigma},q) \bigg]
\eea
where
\bea
V(H,{\bar\sigma},q)=  -M^2_{\rm eff} |H|^2 +\Big(\lambda+\frac{1}{2}c^2_2\Big)  |H|^4 +\Lambda_{\rm eff} +\frac{1}{2} m^2_{\bar\sigma} \Big( {\bar\sigma}-  \sqrt{\frac{3}{2}} \,c_1(c_2 |H|^2+q) \Big)^2
\eea
with
\bea
m_{\bar\sigma}= \sqrt{\frac{2}{3}}\, \frac{M_P}{\sqrt{\zeta^2-c^2_1}}. \label{inflatonmass}
\eea
Thus, in the minimum of the sigma field potential, we get the Higgs potential as in the case with the four-form coupling to the Higgs field only \cite{Giudice,Kaloper}.
We note that the coupling between the sigma and Higgs fields is  of the form, 
$ \frac{c_1c_2m^2_{\bar\sigma} }{M_P} \,{\bar\sigma} |H|^2$, which determines the reheating temperature after inflation. 

For general field values of $\sigma$,  the canonical sigma field ${\bar\sigma}$ in Einstein frame is redefined by
\bea
\sigma = \frac{1}{c_1} \Big(e^{\sqrt{\frac{2}{3}} {\bar\sigma}}-1 \Big),
\eea
and the Einstein frame Lagrangian becomes
\bea
{\cal L}_E 
=\sqrt{-g_E} \bigg[ \frac{1}{2} R(g_E) -\frac{1}{2}(\partial_\mu{\bar\sigma})^2 - e^{-\sqrt{\frac{2}{3}}{\bar\sigma}}\, |D_\mu H|^2- V_E(H,{\bar\sigma}) \bigg]
\eea
with
\bea
V_E(H,{\bar\sigma})&=& \Lambda_{\rm eff}\, e^{-2\sqrt{\frac{2}{3}}{\bar\sigma}}+\frac{3}{4} m^2_{\bar\sigma} \bigg(1-(1+c_1 q)e^{-\sqrt{\frac{2}{3}}{\bar\sigma}}-c_1 c_2\, e^{-\sqrt{\frac{2}{3}}{\bar\sigma}} |H|^2  \bigg)^2 \nonumber \\
&&+ e^{-2\sqrt{\frac{2}{3}}{\bar\sigma}} \Big( -M^2_{\rm eff}|H|^2+  \lambda_{H,{\rm eff}}|H|^4\Big).
\eea
Here, assuming that the SM Higgs is stabilized at $\langle H\rangle=v/\sqrt{2}$ in each dS phase, we can rewrite the above sigma field potential as
\bea
V_E({\bar\sigma}) = V_0(q) + \bigg[\frac{3}{4}m^2_{\bar\sigma}\Big(1+c_1\Big(q+\frac{1}{2}c_2 v^2\Big)\Big)^2+\Lambda_{\rm eff}  \bigg] \Big(e^{-\sqrt{\frac{2}{3}}{\bar\sigma}}- e^{-\sqrt{\frac{2}{3}}{\bar\sigma}_{\rm m}(q)}\Big)^2 \label{finpot}
\eea
where
\bea
e^{-\sqrt{\frac{2}{3}}{\bar\sigma}_{\rm m}(q)}&=& \frac{3m^2_{\bar\sigma}(1+c_1 (q+\frac{1}{2}c_2 v^2))}{3m^2_{\bar\sigma}(1+c_1 (q+\frac{1}{2}c_2 v^2))^2+4\Lambda_{\rm eff}}, \label{minq} \\
V_0(q)&=&\frac{3m^2_{\bar\sigma} \Lambda_{\rm eff}}{3m^2_{\bar\sigma}(1+c_1 (q+\frac{1}{2}c_2 v^2))^2 +4\Lambda_{\rm eff}}. \label{minV}
\eea
Here, we note that the effect of the effective cosmological constant $\Lambda_{\rm eff}$ in Jordan frame is crucial in determining the minimum of the sigma field potential. This is important for a large shift in the minimum of the potential after the membrane nucleation.

\section{Reheating}

Now we discuss the role of the sigma field potential for reheating during or just after the last membrane nucleation. We keep $\langle H\rangle=0$ during the scanning with the flux parameter and regard the sigma field as the inflaton.

\subsection{Reheating during the last membrane nucleation}

We first consider the possibility of reheating during the last membrane nucleation. 
To this, imposing $m_{\bar\sigma}\sim H$, we can allow for the sigma field to start rolling from the initial misalignment after the next-to-last membrane nucleation, that is, the transition from $q=q_c+e$ to $q=q_c$. Then, the sigma field can decay into the SM particles through the Higgs coupling and reheat the Universe. Moreover, as discussed in the previous section, we assume that the last dS phase decays within the Hubble spacetime volume during the last dS phase, that is,  $\gamma<H^4$, in order not to dilute much the radiation produced from the sigma field decay. 

Just before the next-to-last nucleation, we need $q=q_c+e$ and $v=0$, for which eqs.~(\ref{minq}) and (\ref{minV}) become
\bea
e^{-\sqrt{\frac{2}{3}}{\bar\sigma}_{\rm m}(q_c+e)} &\approx &\frac{1}{1+c_1q_c} \Big(1+\frac{8eq_c}{3m^2_{\bar\sigma}(1+c_1 q_c)^2} \Big)^{-1},  \label{min1}\\
V_0 (q_c+e) &\approx & \frac{6m^2_{\bar\sigma} e q_c}{3m^2_{\bar\sigma}(1+c_1 q_c)^2 +8e q_c}
\eea
where we used $\Lambda_{\rm eff}(q_c-e)=\Lambda+\frac{1}{2}(q_c-e)^2\simeq 0$ in the end, and
\bea
\Lambda_{\rm eff}(q_c+e)= \Lambda + \frac{1}{2} (q_c+e)^2 \simeq 2e q_c.
\eea
On the other hand, after the next-to-last nucleation, we have  $q=q_c$ and $v=0$, for which 
\bea
e^{-\sqrt{\frac{2}{3}}{\bar\sigma}_{\rm m}(q_c)} &\approx &\frac{1}{1+c_1 q_c} \Big(1+\frac{4eq_c}{3m^2_{\bar\sigma}(1+c_1 q_c)^2} \Big)^{-1},  \label{min1}\\
V_0 (q_c) &\approx & \frac{3m^2_{\bar\sigma} e q_c}{3m^2_{\bar\sigma}(1+c_1 q_c)^2 +4e q_c}
\eea
where use is made of
\bea
\Lambda_{\rm eff}(q_c)= \Lambda + \frac{1}{2} q^2_c = e\Big(q_c-\frac{1}{2}e\Big)\approx e q_c.
\eea
Thus, we find that both the minimum of the sigma field potential and the cosmological constant changes after the next-to-last nucleation. 

Taking the initial condition just before the next-to-last nucleation to be the minimum of the potential for $q=q_c+e$, i.e. ${\bar\sigma}_i={\bar\sigma}_{\rm m}(q_c+e)$, we can obtain the sigma field potential after the next-to-last nucleation as
\bea
V_E({\bar\sigma})
&\approx&V_0(q_c+e) \nonumber 
\\&&+ \frac{1}{4}[3m^2_{\bar\sigma}(1+c_1 q_c)^2+8 eq_c] e^{-2\sqrt{\frac{2}{3}}{\bar\sigma}_{\rm m}(q_c+e)} 
 \Big(e^{-\sqrt{\frac{2}{3}}({\bar\sigma}-{\bar\sigma}_{\rm m}(q_c+e))}- e^{-\sqrt{\frac{2}{3}}({\bar\sigma}_{\rm m}(q_c)-{\bar\sigma}_{\rm m}(q_c+e))}\Big)^2 \nonumber \\
 &=&V_0(q_c+e) \nonumber \\
 &&+ \frac{3}{4}m^2_{\bar\sigma}\bigg(1+\frac{8 eq_c}{3m^2_{\bar\sigma} (1+c_1 q_c)^2} \bigg)^{-1} \Big(e^{-\sqrt{\frac{2}{3}}({\bar\sigma}-{\bar\sigma}_i)}-1-\frac{4eq_c}{3m^2_{\bar\sigma} (1+c_1 q_c)^2+4 e q_c}\Big)^2
\eea
with
\bea
V_0(q_c+e)\simeq \frac{6m^2_{\bar\sigma}eq_c}{3m^2_{\bar\sigma}(1+c_1 q_c)^2+8e q_c}.
\eea
As a result, the sigma field starts to oscillate at ${\bar\sigma}={\bar\sigma}_i$ with the initial potential energy, given by
\bea
V_i&\equiv& V_0(q_c+e) + \frac{36 m^4_{\bar\sigma}(eq_c)^2}{[3m^2_{\bar\sigma}(1+c_1 q_c)^2+8e q_c][3m^2_{\bar\sigma}(1+c_1 q_c)^2+4e q_c]^2} \nonumber \\
&=&\frac{6m^2_{\bar\sigma}eq_c}{3m^2_{\bar\sigma}(1+c_1 q_c)^2+8e q_c} \bigg(1+\frac{6m^2_{\bar\sigma}eq_c}{[3m^2_{\bar\sigma}(1+c_1 q_c)^2+4e q_c]^2} \bigg).
\label{maxpot0}
\eea

Then, the sigma field starts to oscillate around the minimum of the above potential, provided that the Hubble parameter at $q=q_c$ satisfies $H(q_c)= m_{\bar\sigma,{\rm eff}}$, i.e.
\bea
H(q_c)= \frac{V_i}{\sqrt{3}}=m_{\bar\sigma,{\rm eff}} \label{inflatonmass}
\eea
where
\bea
m^2_{\bar\sigma,{\rm eff}} \equiv \frac{3m^4_{\bar\sigma}(1+c_1 q_c)^2}{3m^2_{\bar\sigma}(1+c_1 q_c)^2+8e q_c}.
\eea
This implies that we need $m^2_{\bar\sigma}\simeq 2 eq_c$.
For $q>q_c$, the Hubble parameter $H(q)$ becomes larger than the sigma field mass so the sigma field is stuck at a certain value or it undergoes a slow rolling. Therefore, only when the sigma field mass is appropriately chosen for a given flux parameter, the sigma field can start to oscillate and reheat the Universe at $q=q_c$. After the final membrane nucleation at $q=q_c-e$, the Higgs mass parameter becomes negative and takes a right value for the observed Higgs mass and the cosmological constant also takes the observed value by the anthropic argument. 

However, the sigma field couples to the SM Higgs through the non-minimal coupling to the four-form flux, which is suppressed by the Planck scale.
Consequently,  from the decay rate of the inflaton for the decay into two Higgs fields as
\bea
\Gamma_{\bar\sigma}= \frac{3c^2_1 c^2_2}{64\pi}  \frac{m^3_{\bar\sigma}}{M^2_P}, \label{decay}
\eea
and using eq.~(\ref{inflatonmass}), we obtain the reheating temperature as
\bea
T_{\rm RH} &=&\left(\frac{90}{\pi^2 g_*} \right)^{1/4} (M_P \Gamma_{\bar\sigma})^{1/2} \nonumber \\
&=&0.2\left(\frac{100} {g_*} \right)^{1/4} c_1 c_2\, \bigg(\frac{e}{c_2 M^{2/3}} \bigg)^{3/4} \bigg(\frac{M}{M_P} \bigg)^2.
\eea
For instance, in order to solve the hierarchy between the Planck scale and the weak scale by the relaxation of the four-form flux, we choose $M\sim M_P$ and $\sqrt{e}\sim 1\,{\rm TeV}$ for $c_2={\cal O}(1)$. Then, the inflaton mass is $m_{\bar\sigma}\sim {\rm TeV}$ and the reheating temperature is $T_{\rm RH}\sim (c_1/10^3)\,10\,{\rm MeV}$. In other words, we need $\zeta\sim 10^{15}$ for  $m_{\bar\sigma}\sim {\rm TeV}$, and $c_1\gtrsim 10^3$ for $T_{\rm RH}\gtrsim 10\, {\rm MeV}$. 

We comment on several issues in the case with reheating during the last membrane nucleation.
First, after the next-to-last membrane nucleation, it is known that the Universe enters the open inflating phase with a negative spatial curvature \cite{coleman,hawking}.
If the tunneling with the last membrane nucleation were efficient, there would no time for the negative spatial curvature to be diluted away by inflation, so it remains sizable after the last membrane nucleation. 

Secondly, we would need a large coupling of quadratic curvature gravity for a light dual scalar field. In this case, certainly, the perturbative expansion of the tree-level Lagrangian is in question. Moreover, we need to control even higher curvature terms such as $R^n$ for $n>2$ with sufficiently small coefficients. In this sense, there is a need of improving our discussion to justify the classical Lagrangian for the light sigma field at the $R+R^2$ gravity level. At least, we can argue that when we take the pure $R+R^2$ gravity as the effective theory, the UV cutoff scale for gravity does not decrease, because the theory is identical to a scalar-tensor gravity with a stable massive scalar field.

In the next subsection, we also discuss a successful case of reheating after the last membrane nucleation but without problems of the open Universe or large couplings of quadratic curvature gravity or four-form flux.

\subsection{Reheating after the last membrane nucleation}

In the case when the last membrane nucleation takes a longer than the Hubble rate, that is, $\gamma<H^4$, radiation produced during the last dS phase would be diluted away by the exponential expansion of the Universe. 
Thus, in this subsection, we discuss the case when reheating occurs after the last membrane nucleation.

Just after the last nucleation, we have $q=q_c-e$, $v\neq 0$ and $V_0\approx 0$, for which the minimum of the potential becomes
\bea
e^{-\sqrt{\frac{2}{3}}{\bar\sigma}_{\rm m}(q_c-e)}\approx \frac{1}{1+c_1 (q_c-e+\frac{1}{2}c_2 v^2)}\approx\frac{1}{1+c_1 q_c}. \label{min2}
\eea
Then, we can compare between the different minimum values in eqs.~(\ref{min1}) and (\ref{min2}) before and after the last membrane nucleation, which are crucial for obtaining a nonzero initial vacuum energy for the sigma field after the last membrane nucleation.  The discussion on the flux-induced shift of the minimum and reheating has been generalized to the case with the four-form couplings to singlet scalar fields in a recent paper \cite{general}. 

Suppose that the sigma field settles into the minimum of the potential before the last nucleation.
Then, after the last nucleation, the minimum of the potential is shifted from eq.~(\ref{min1}) to eq.~(\ref{min2}). Taking the initial condition just before the last nucleation to be the minimum of the potential for $q=q_c$, i.e. ${\bar\sigma}'_i={\bar\sigma}_{\rm m}(q_c)$, we can obtain the sigma field potential after the last nucleation as
\bea
V_E({\bar \sigma})
&\approx&\frac{3}{4}m^2_{\bar\sigma}(1+c_1 q_c)^2 e^{-2\sqrt{\frac{2}{3}}{\bar\sigma}_{\rm m}(q_c)} 
 \Big(e^{-\sqrt{\frac{2}{3}}({\bar\sigma}-{\bar\sigma}_{\rm m}(q_c))}- e^{-\sqrt{\frac{2}{3}}({\bar\sigma}_{\rm m}(q_c-e)-{\bar\sigma}_{\rm m}(q_c))}\Big)^2 \nonumber \\
 &=& \frac{3}{4}m^2_{\bar\sigma} \Big(1+\frac{4eq_c}{3m^2_{\bar\sigma}(1+c_1 q_c)^2} \Big)^{-2} \Big(e^{-\sqrt{\frac{2}{3}}({\bar\sigma}-{\bar\sigma}'_i)}-1-\frac{4eq_c}{3m^2_{\bar\sigma}(1+c_1 q_c)^2} \Big)^2.
\eea
As a result, the sigma field starts to oscillate at ${\bar\sigma}={\bar\sigma}_i$ with the initial potential energy, given by
\bea
V'_i\equiv V_E({\bar\sigma}_i) =\frac{12(e q_c)^2 m^2_{\bar\sigma}}{(3 m^2_{\bar\sigma} (1+ c_1q_c)^2+4 eq_c )^2} \label{maxpot}
\eea
where the latter approximation is made for $c_1 q_c\lesssim 1$.
Here, we find that: for $m^2_{\bar\sigma}\ll eq_c$,  $V'_i\approx \frac{3}{4}  m^2_{\bar\sigma}$; for $  m^2_{\bar\sigma}\gg eq_c$, $V'_i\approx  \frac{4}{3} (e q_c)^2/[m^2_{\bar\sigma}(1+c_1 q_c)^2]$.
On the other hand, for $m^2_{\bar\sigma}=\frac{2}{3}\sqrt{2} eq_c/(1+c_1 q_c)^2 $, the initial potential energy is maximized to $V'_i\approx 0.25(e q_c)/(1+c_1 q_c)^2$.
Thus, the maximum initial potential can be obtained for the inflaton mass of order $1\,{\rm TeV}$  for $e\sim(1\,{\rm TeV})^2$ and $q_c\sim M^2_P$, but a heavier inflaton mass is favored for a sufficiently high reheating temperature as will be shown below.

When reheating is instantaneous, the temperature of the Universe after inflation would be given by the maximum temperature, $T_{\rm max}=\left( \frac{90 V'_i}{\pi^2 g_*} \right)^{1/4}$ with eq.~(\ref{maxpot}),  thus becoming
\bea
T_{\rm max}&\simeq& 2.5\times 10^{10}\,{\rm GeV} \left(\frac{100}{ g_*} \right)^{1/4}  \left(\frac{eq_c}{(1\,{\rm TeV}\cdot M_P)^2}\right)^{1/4}  \nonumber \\
&&\quad\times \bigg(\frac{m^2_{\bar\sigma} M^2_P}{eq_c}\bigg)^{1/4}\left(1+\frac{3}{4}\left(\frac{m^2_{\bar\sigma}M^2_P}{eq_c}\right)(1+c_1 q_c/M^2_P)^2 \right)^{-1/2} 
\eea 
where we have reintroduced the Planck scale for dimensionality. 
In particular, for $m^2_{\bar\sigma}\gg eq_c$ and $c_1 q_c/M^2_P\lesssim 1$, the maximum reheating temperature becomes
\bea
T_{\rm max}&\simeq& 1.5\times 10^{9}\,{\rm GeV} \left(\frac{100}{ g_*} \right)^{1/4}  \left(\frac{eq_c}{(1\,{\rm TeV}\cdot M_P)^2}\right)^{1/2}  \left(\frac{380\,{\rm TeV}}{m_{\bar\sigma}}\right)^{1/2}.
\eea
Then, from the decay rate of the sigma field given in eq.~(\ref{decay}), the resulting reheating temperature becomes much lower than the maximum reheating temperature, as follows, 
\bea
T_{\rm RH} =\left(\frac{90}{\pi^2 g_*} \right)^{1/4} ( \Gamma_{\bar\sigma} M_P)^{1/2}=  10\,{\rm MeV}\left(\frac{100}{ g_*} \right)^{1/4} \Big(\frac{c_1}{1} \Big) \Big(\frac{c_2}{1} \Big) \Big(\frac{m_{\bar\sigma}}{380{\rm TeV}}\Big)^{3/2}.
\eea
In this case, the reheating temperature is much smaller than the maximum temperature, due to the double suppressions with the Planck scale and the inflaton mass. However, for $m_{\bar\sigma}>380\,{\rm TeV}$ (or $\zeta<5.2\times 10^{12}$ from eq.~(\ref{inflatonmass})) and $c_1,c_2={\cal O}(1)$, we can obtain a sufficiently high reheating temperature for the successful Big Bang Nucleosynthesis. 
We note that for $m_{\bar\sigma}\geq 1.6\times 10^8\,{\rm GeV}$, the reheating temperature becomes identical to the maximum reheating temperature,  that is, $T_{\rm RH}=T_{\rm max}$. 

In the discussion of this subsection, we showed that there is no need of large couplings to the four-form flux for the successful reheating.
But, we still need to see the details of the sufficient number of efoldings and the inflationary observables in a low-scale inflation.

\section{Conclusions}

We provided the most general Lagrangian for the four-form couplings to the SM and showed that the  four-form flux parameter scans not only the Higgs mass and the cosmological constant but also the Planck mass.  We found that the non-minimal four-form coupling to gravity gives rise to an tachyonic instability for a new scalar field, but it can be consistently cured in the effective Lagrangian. We discussed the conditions on new four-form couplings for a successful reheating of the Universe at the end of relaxation.

\section*{Acknowledgments}

The author thanks Kfir Blum, Cliff Burgess, Gian Giudice, Jinn-Ouk Gong and Kimyeong Lee for helpful discussions. 
The author also appreciates fruitful discussions with participants during the CERN-CKC Theory Workshop on Axion in the lab and in the cosmos, and attendees in the BSM forum in CERN theory group in October 2019. The important comments from the anonymous referee of the journal also helped much improving the discussion of the paper. 
The work of HML is supported in part by Basic Science Research Program through the National Research Foundation of Korea (NRF) funded by the Ministry of Education, Science and Technology (NRF-2019R1A2C2003738 and NRF-2018R1A4A1025334).




\end{document}